# Nanoparticles' Passive Targeting Allows Optical Imaging of Bone Diseases

*Chao Mi[1,]\*, Xun Zhang[1], Chengyu Yang[2], Jianqun Wu[5], Xinxin Chen[3], Chenguang Ma[1], Sitong Wu[1,4], Zhichao Yang[1,4], Pengzhen Qiao[1], Yang Liu[2], Weijie Wu[1], Zhiyong Guo[1,5], Jiayan Liao[4], Jiajia Zhou[4], Ming Guan[1], Chao Liang[3,]\*, Chao Liu[2,5,]\*, and Dayong Jin[1,4,]\**

1. UTS-SUSTech Joint Research Centre for Biomedical Materials and Devices, Department of Biomedical Engineering, Southern University of Science and Technology, Shenzhen, China

2. Department of Biomedical Engineering, Southern University of Science and Technology, Shenzhen, China

3. Department of Biology, School of Life Sciences, Southern University of Science and Technology, Shenzhen, China

4. Institute for Biomedical Materials and Devices (IBMD), Faculty of Science, University of Technology Sydney, Australia

5. Guangdong Provincial Key Laboratory of Advanced Biomaterials, Southern University of Science and Technology, Shenzhen, China

*Corresponding author (email to: jindy@sustech.edu.cn (D. Jin); liuc33@sustech.edu.cn (C. Liu); liangc@sustech.edu.cn (C. Liang); mic@sustech.edu.cn (C. Mi))

**Abstract:**

Bone health related skeletal disorders are commonly diagnosed by X-ray imaging, but the radiation limits its use. Light excitation and optical imaging through the near-infrared-II window (NIR-II, 1000-1700 nm) can penetrate deep tissues without radiation risk, but the targeting of contrast agent is non-specific. Here, we report that lanthanide-doped nanocrystals can be passively transported by endothelial cells and macrophages from the blood vessels into bone marrow microenvironment. We found that this passive targeting scheme can be effective for longer than two months. We therefore developed an intravital 3D and high-resolution planar imaging instrumentation for bone disease diagnosis. We demonstrated the regular monitoring of 1 mm bone defects for over 10 days, with resolution similar to X-ray imaging result, but more flexible use in prognosis. Moreover, the passive targeting can be used to reveal the early onset inflammation at the joints as the synovitis in the early stage of rheumatoid arthritis. Furthermore, the proposed method is comparable to μCT in recognizing symptoms of osteoarthritis, including the mild hyperostosis in femur which is ~100 μm thicker than normal, and the growth of millimeter-scale osteophyte in the knee joint, which further proves the power and universality of our approach in diagnosis of bone diseases.

**Significance**

This work opens a new direction for non-invasive diagnosis of bone diseases by the NIR-II imaging with high spatial resolution. The purpose-engineered molecular sized nanoparticles enable the ultrasensitive in vivo 3D and planar skeleton imaging, meanwhile the interrelated bone marrow cell uptake and the body clearance have been revealed. This yields not only the long-term monitoring of the living mice for more than two months, but also the accurate diagnosis on a range of common bone diseases, including the cortical bone defect (1 mm in diameter), synovitis, rheumatoid arthritis, osteoarthritis, osteophyte, and hyperostosis.

Mammalian bone, together with tendons and muscles, perform the vital functions of supporting the body weight and locomotion. A microscopic view of the bone structure reveals its extensive blood vessel networks and the niche for both the mesenchymal and hematopoietic stem cells, as well as their progenies[1]. Bone health is critical in maintaining body movement capability, immunity and metabolism[2]. However, a series of commonly occurred bone diseases, including fracture, skeletal deformity, osteoporosis, rheumatoid arthritis, osteoarthritis, bone tumor, are hard to become noticed until the late-stage symptoms appear, unless X-ray imaging can be use[3–6]. Nonetheless, the regular exposure to X-radiation will cause DNA damage and leukocyte death and is classified as a "known human carcinogen" by World Health Organization. More specifically, a study in Australia shows that exposure to an X-ray computed tomography (CT) scan (average effective radiation dose 4.5 mSv) in childhood or adolescence leads to a 24% higher overall cancer incidence, and the incidence rate ratio increases with each additional CT scan[7]. According to the US Food and Drug Administration, 10 mSv, the effective dose of radiation from a CT scan of the abdomen and pelvis, might increase the risk of cancer by about 1 in 2000[8]. To date, there is no other alternative imaging approach for visualizing the microscopic structure and monitoring the dysfunctions of bone health *in vivo*.

Optical imaging allows regular and real-time visualization of cells and their functions [9–11]. More significantly, compared with the visible (400-700 nm) and NIR-I (700-900 nm) ranges, light excitation and emission at the NIR-II window (1000-1700 nm) can penetrate deep tissues, which allows high spatial and temporal resolutions to be achieved with high signal-to-background ratio (SBR), as the long wavelength of light leads to minimal tissue scatterings and autofluorescence[12–15]. Towards the realization of the above potentials, NIR-II emitting fluorescent materials, including semiconductor quantum dots (QDs)[16–18], carbon nanotubes[19,20], lanthanide doped nanocrystals (LnNCs)[21–25], and organic molecules[26–30], have been developed for theranostic applications[31–33]. However, the efficient delivery of these nanoparticles to target the cellular structure of the bone remains challenging as the nanoparticles are typically filtered by the mononuclear phagocyte system, especially in the liver and spleen[34,35], not mentioning the slow blood flow and the high density of bone composites[36]. These cascade barriers made the use of positive targeting ligands ended up with marginal improvement to approximately 0.9% of injected dose[37,38].

Here, we report that a set of lanthanide doped nanoparticles emitting at ~1550 nm can passively migrate into bone marrows from the bloodstream by spontaneous cellular transport. The nanoparticles show high permeability inside the skeleton microenvironment and as NIR-II contrast agents allow the long-term high-resolution imaging of early onset of bone diseases. We demonstrate the biomedical translation potential for accurate diagnosis of skeleton diseases, including cortical bone defect, osteophyte, hyperostosis, osteoarthritis, synovitis, and rheumatoid arthritis.

## Results and discussions

**Preparation of ErNCs NIR-II probe.** We first carefully synthesized the $NaYbF_4$: $Er^{3+}$, $Ce^{3+}$ with the high degree of morphological and optical uniformity. We then proceeded with heteroepitaxial growth of an inert shell to greatly enhance the fluorescence by passivating surface quenchers, yielding the high performance $NaYbF_4$: $Er^{3+}$, $Ce^{3+}$@$NaYF_4$ (ErNCs) NIR-II probe, as shown in Figure 1a. We than grafted ErNCs with PEG (polyethylene glycol) for prolonging blood circulation and reducing hepatic uptake (Figure 1a), as the administrated PEG-coated ErNCs can escape from mononuclear phagocytes in liver and spleen. The prepared nanocrystals at each step showed a high degree of uniform morphology and dispersibility (Figure 1b), with a NIR-II emission band of the PEG-coated ErNCs around 1550 nm (Figure 1c). According to cellular toxicity assays, shown in Figure S1, PEG-coated ErNCs displayed a high degree of biocompatibility.

**Selection of ErNCs as the NIR-II contrast agent for high-resolution imaging.** We further confirmed that the ~1550 nm NIR-IIb (1500-1700 nm) emission of ErNCs can largely enhance the imaging resolution, compared with the emissions from NIR-IIa (1000-1500 nm) fluorophores[11,39], such as the ~1000 nm emission of ICG, the ~1060 nm emission of $NaYF_4$: $Nd^{3+}$ (NdNCs), and the ~1340 nm emission of NdNCs, as shown in Figures 1d and S2. With a standard *ex vivo* test, Figures 1e and 1f further quantitatively displayed the resolution enhancements by a full width at half maximum (FWHM) as 1.05, 0.86, 0.78 and 0.64 mm for the 1000, 1060, 1340, and 1550 nm channel, respectively, and the *in vivo* results on the lymph vessels imaging confirmed that ErNCs with longer emission wavelength are superior to NdNCs (Figures 1g and 1h).

**Long-term *in vivo* imaging of bones for over two months.** After the tail-vein injection of ErNCs for ~4 hours, the NIR-II signals from the blood vessel gradually vanished (Figures S3 and S4), suggesting the fast clearance of nanoparticles from blood circulation. Interestingly, NIR-II imaging revealed that the nanocrystals were gradually accumulated in bones. The bone structures became clearly visualized with high resolution and contrast both *in vivo* and *ex vivo*. Three days after the tail-vein injection, the dorsal, chest and limb skeleton structures, including the skull, spine, sternum, tibia, phalanx, rib, and femur, have been clearly resolved, as the *in vivo* images shows in Figure 2a. Moreover, morphology details, including multiple segments in the spine and sternum, the growth plate, and the subchondral bone (the zoom-in area at the knee joint, confirmed by μCT in Figure S5), can be clearly visualized. In contrast, the results achieved by NdNCs were ambiguous with a lower resolution (Figure 2b). Remarkably, long-term *in vivo* imaging of bone structures has been achieved for more than 30 days (Figure S6). The NIR-II signal from bones continued to increase up to 5 days after the injection (Figures 2c and 2d) and eventually disappeared after a prolonged observation timepoint of over 70 days (Figure S6).

**Three-dimensional (3D) *in vivo* and *ex vivo* imaging.** 3D imaging could provide more realistic and abundant morphological characteristics than two-dimensional patterns. To develop the 3D bone imaging, we utilized a multi-angle rotation method to construct the multi-view images of the mouse tibia. As shown in Figure 2e, the reconstructed stereo image vividly shows the 3D structure of the tibia and the knee joint. Moreover, the *ex vivo* 3D imaging result matches well with the *in vivo* result (Figure 2e, Video 1 and 2). Compared with X-ray and CT imaging, the NIR-II imaging approach is radiation free, and allows for the long-term, non-invasive, and time-critical diagnosis of skeletal disorders associated with bone diseases.

**Passive targeting of nanoparticles to bone marrow.** The *in vivo* transportation and orientation of nanoparticles has profound significance on its biomedical applications, including bone-targeted imaging and drug delivery[5]. Recent research argues that the passive-targeting is mainly due to the unique hydroxyapatite mineral binding ability of PEG-coated nanoparticles[40]. In our study, the fluorescence from bone vanished after flushing out the bone marrow, while marrow mesenchyme showed a strong signal (Figure 3a). Furthermore, after decalcification, the NIR-II signal was dramatically improved compared to the undecalcified bones (Figure 3b). These results confirmed that ErNCs mainly accumulate in bone marrow, rather than bonding with the cortical bone.

This motivated us to investigate deeply into the bone marrow targeting mechanism. By co-localization confocal imaging on marrow sections, we confirmed that the Cy3 fluorescence of ErNCs@Cy3 overlaps with the channel of macrophages (Figure 3c) with a Pearson correlation coefficient of up to 0.57, which suggests the macrophage's uptake of ErNCs. This is because plasma proteins (e.g., immunoglobulins, adhesion mediators, complement proteins) may act as opsonins to bind with the nanomaterials, leading to efficient phagocytosis by macrophages[35,41,42]. Considering the slow blood flow in the dense capillary network inside marrows[36,43], the local macrophages take their time to capture nanoparticles and transport them to a distance from the vessels[44].

Studies suggested that up to 97% of nanoparticles enter tumours through endothelial cells in blood vessels[45]. We hypothesise a similar process in marrows. The fluorescence co-localization results verified the endocytosis of ErNCs by endothelial cells (Figure 3d). Conversely, the whole area of the cortical bone shows no Cy3 signal at all (Figure 3e). In conclusion, with long blood circulation time, ErNCs' passive transportation by endothelial cells and macrophages in marrows underpin the remarkable bone targeting ability.

**Toxicity studies and intact nanocrystals' body clearance.** Next, we carefully examined the potential toxicity that may be caused by the accumulation of ErNCs in bone marrow. As shown in Figure S7, the H&E staining tissue sections show no apparent damage or abnormality on cellular structures after 14 and 70 days post injection of ErNCs. Moreover, we monitored the stable NIR-II fluorescence from feces for 13 days after the injection, suggesting that the nanoparticles were gradually excreted (Figure S8). The prepared ErNCs for *in vivo* NIR-II imaging is biocompatible, metabolizable, and safe to use.

***In vivo* NIR-II detection of bone defect for up to 11 days.** To evaluate the imaging resolution, we purposely operated the ~1 mm mono-cortical bone defects on the mice tibias (Figure 4a)[3,46]. The X-ray imaging proves the injured bone sites with around 1 mm diameter (Figure 4b). Impressively, the *in vivo* NIR-II imaging diagnoses also showed a similar result of 1.2 mm in cortical bone defect diameter, albeit the NIR-II intensity decreases after the surgery (Figures 4a and 4b). Notably, the fluorescent intensity drop within 6 mm around the injured site could be caused by impaired blood circulation.

As the surgery causes collateral damage to the soft tissue and induces the poor blood circulation of ErNCs on the surgical site, we postponed the injection of ErNCs by 5 days after the bone surgery to allow the damaged vascular network to heal. As shown in Figure 4c, long-term monitoring from Day 6 to 11 after the bone defect surgery showed a shaded area that matched the size of the bone defect. The control group with the sham surgery showed slight abnormity around the soft tissue wound on Day 6 after surgery. The cross-sectional intensity further confirms the size of the bone defect on the left hindlimb as 1.2 mm (Figure 4d), compared with the result of no defect being detected on the right hindlimb 11 days after surgery, as the bone injury recovery is usually slower than soft tissue wound healing (Figure S9). More cases of NIR-II imaging, shown in Figure S10, accurately resolved the sizes of the bone defects as 1.2, 1.36 and 1 mm, arbitrarily made in surgery.

**High-specificity-inflammation imaging to reveal the early onset and progression of rheumatoid arthritis (RA).** RA is a chronic autoimmune skeleton disease with a worldwide annual incidence of 3 cases per 10,000 and a prevalence rate of 1%[47,48]. The NIR-II imaging opens the new opportunity for RA diagnosis. NIR-II imaging in collagen-induced arthritis (CIA), the gold standard animal model for RA, illustrated that two toes displayed enhanced NIR-II intensity compared with the other toes from one mouse paw with mild RA, while the heavily swollen paw caused by severe RA was overexposed in all the toe joints, in sharp contrast to the clearly visualized bones and joints of the normal mouse paw (Figures 5a and S11). The μCT analysis consistently showed the aggravation of RA (Figures 5b-5d), as evidenced with 3D images of paws and bone mineral density (BMD) and a microarchitecture parameter of bone (BV/TV), which proves the potential of NIR-II imaging for monitoring RA progression.

**Accurate recognition of synovitis.** Early diagnosis and treatment of RA are limited by its unknown etiology and initial similarity to other inflammatory diseases[49]. The pathophysiology of RA starts with chronic inflammation of the synovial membrane, then erosion of articular cartilage and juxta-articular bone[48]. If untreated, chronic RA can lead to systemic inflammation resulting in abnormalities in heart, liver, and other organs[47]. Thus, early diagnosis of RA depends on the accurate recognition of synovitis. As shown in Figures 5e and S12a, after the injection of ErNCs, the toes with inflammation exhibit higher intensities because of the active phagocytosis of ErNCs by proliferative macrophages. Significantly, such localized intensity enhancement has high specificity and sensitivity to inflammation, because only the distal joint with higher NIR-II signal is indeed inflamed but not the proximal one with normal intensity from the same toe, which is confirmed by immunofluorescence imaging of inflammatory factor IL-1β (Figures 5f and S12b). Moreover, with high resolution, the enhanced intensity pattern matched well with the distribution of synovium on the side of the joint (Figures 5g and S12c), which excludes similar inflammatory diseases and determines as synovitis.

**NIR-II imaging diagnosis of osteoarthritis.** Osteoarthritis is the most common degenerative joint disease and the leading cause of physical disability. It occurs with the formation of osteophytes and cartilage loss[4,50], as illustrated in Figure 6a. To demonstrate the power of NIR-II imaging in diagnosing osteoarthritis, we established a mouse model induced by anterior cruciate ligament transection (ACLT) surgery. The μCT and safranin O-fast green staining results (Figure 6b) confirmed the gradual cartilage degeneration and the bone mass decrease during the formation of osteoarthritis. As shown in Figure 6c, while both NIR-II and μCT images showed normal morphology from three angles of view for the right knee joint as the control, the obvious formation of the millimeter-sized osteophyte can be distinctly resolved at the left knee joint after ACLT surgery by both NIR-II and μCT image result.

Interestingly, we also found the drop of local NIR-II intensity could suggest the growth of hyperostosis. As shown in Figure 6d, while the NIR-II images for all three control cases reveal both femur and tibia, the femur structure in the left knee with osteoarthritis cannot be imaged as clearly, since the NIR-II intensity gap Δ2 in the osteoarthritis model is over 2 times higher than Δ1 in the control cases (Figure 6e). These results may indicate the formation of hyperostosis in the cortical bone and the growth of osteophyte as the symptoms of osteoarthritis[4,50], which will block the NIR-II signal from the femur marrow. The μCT examinations confirmed this hyperostosis diagnosis by NIR-II imaging, as the medial condyles of the left femurs showed increased bone thickness over 200 μm, while the right femurs as control were generally thinner than 100 μm (Figures 6f and 6g, and see more comparative trials in Figures S13-S15). Contrastively, it is highly possible that no hyperostosis growth at tibias in the ACLT group, because the NIR-II intensities were similar to the control group. The subsequent μCT measurements proved this conclusion that both ACLT and control groups have normal and similar bone thicknesses at the tibias (Figures S16-S17).

## Conclusion and outlook

While the NIR-II *in vivo* imaging has attracted enormous attentions in the past decade, investigations on the efficiency of the NIR-II fluorescence agents, their bone-targeting mechanisms, body clearance pathway, biocompatibility and toxicity are still in their infancy[51,52]. By developing molecular-scale nanocrystals with high brightness ~1550 nm emission and surface PEGylation, this study has realized noninvasive, high-flexibility and high-resolution imaging diagnosis of small cortical bone defects, rheumatoid arthritis and osteoarthritis. Compared with the conventional X-ray and μCT techniques, NIR-II imaging can provide detailed incidence features and achieve the matchable diagnosis results, but more prospectively, is free of radiation risk for regular use and long-term monitoring of the potential bone disorders and treatment progression. The rapid progress made in high-resolution optical imaging systems and high efficiency optical materials, as well as the ongoing efforts in studying *in vivo* transport and specific targeting of nanoparticles with different sizes and surface conditions with improved body clearance, will continue to advance the field of NIR-II imaging towards pre-clinical and clinical translations.

## Methods

Details and any associated references are provided in the Supplementary Information.

## Data availability

All relevant data that support the findings of this study are available within this published article or available from the corresponding author upon reasonable request.

**Acknowledgments**

The authors acknowledge the financial support from the National Natural Science Foundation of China (62005179, 82172386, 81922081), China Postdoctoral Science Foundation (2020M682866), the Shenzhen Science and Technology Program (KQTD20170810110913065, 20200925174735005), Shenzhen Science and Technology Innovation Commission (KQTD20200820113012029, JCYJ20210324104201005), National Natural Science Foundation of China (62005116, 51720105015), and Guangdong Provincial Key Laboratory of Advanced Biomaterials (2022B1212010003).

**Author contributions**

D. Jin and C. Mi conceived the project. D. Jin, C. Liu, C. Liang and C. Mi supervised the research. X. Zhang, C. Yang, C. Mi, J. Wu, X. Chen and Y. Liu prepared the animal model and performed the imaging. S. Wu, Z. Yang, C. Mi and Z. Guo carried out the 3D imaging. C. Mi, C. Ma, P. Qiao and W. Wu synthesised and characterised the nanocrystals. C. Mi, D. Jin and C. Liu conducted the research experiments and data analysis. C. Mi and D. Jin prepared the figures, supplementary materials and wrote the manuscript with input from other authors. All authors participated in the discussion of the results.

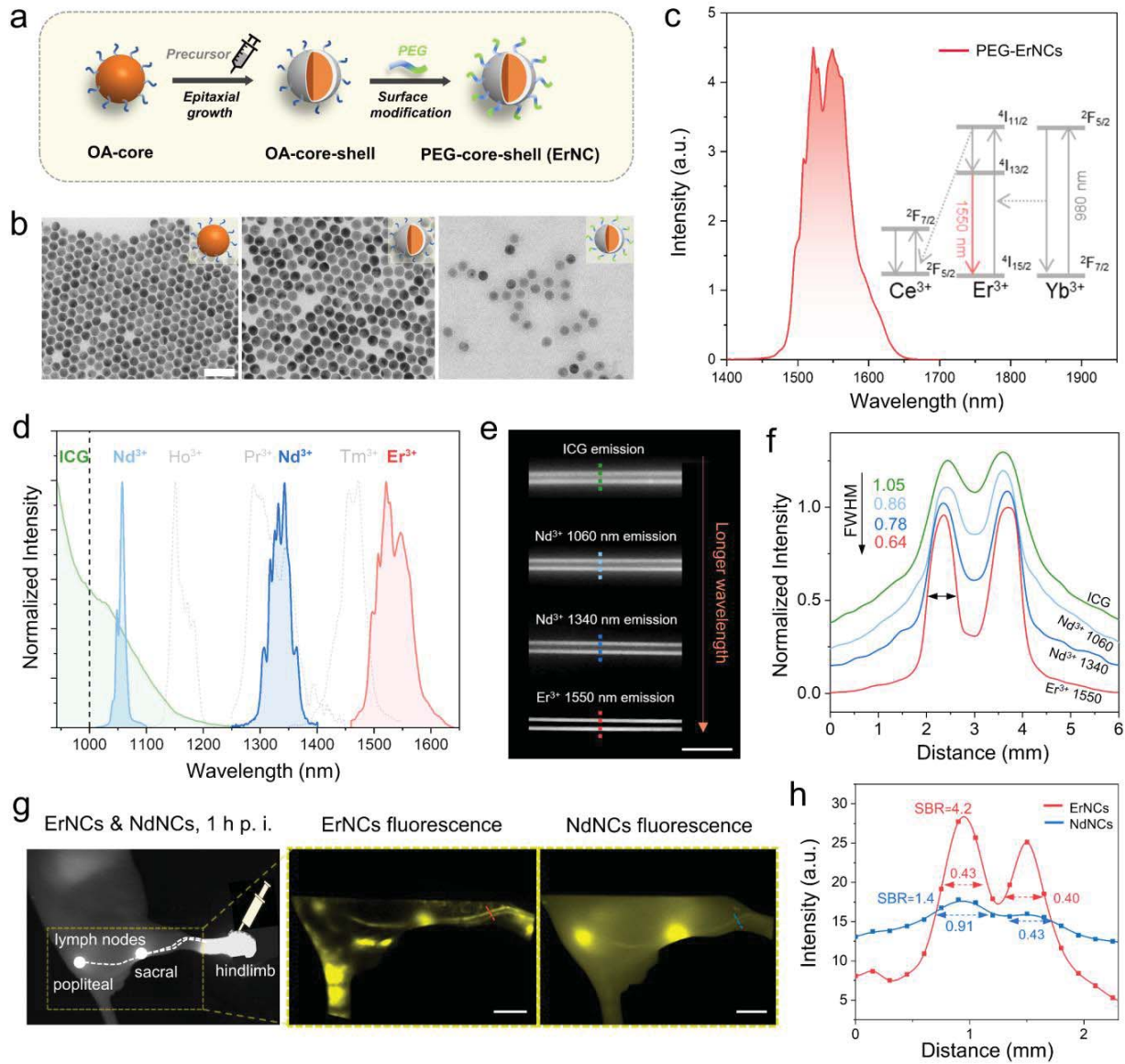

**Figure 1. Design and characterization of ErNCs for high-resolution NIR-II imaging. a**, The preparation of ErNCs. **b**, The TEM results of synthesized OA-capped core and core-shell ErNCs, and PEG-capped core-shell ErNCs from left to right, respectively. The doping concentration is 92% for Yb³⁺, 4% for Er³⁺ and 4% for Ce³⁺. Scale bar: 100 nm. **c**, The emission spectrum of the prepared PEG-capped core-shell structure ErNCs under a 980 nm continuous wave laser excitation. The insert illustrates the simplified energy transfer route among the doped Yb³⁺, Er³⁺ and Ce³⁺ ions for ~1550 nm emission. **d**, The collection of the NIR-II emissions at the different spectrum bandwidths from LnNCs (Ln=Nd³⁺, Ho³⁺, Pr³⁺, Tm³⁺, Er³⁺), and a clinically approved dye ICG. **e**, Representative NIR-II fluorescence images of glass capillaries covered by the 2 mm thick 1% fat emulsion, filled by ICG (~1020 nm long-pass), NdNCs (1064 nm band-pass and ~1340 nm band-pass respectively) and ErNCs (~1319 nm long-pass), respectively. Scale

bar: 5 mm. **f**, Normalized signal intensity profiles across the cross-sections of corresponding fluorescence images indicated by dash lines in (**e**). **g**, *In vivo* NIR-II imaging of lymphatic system by ErNCs (980 nm excitation, 38 mWcm$^{-2}$, 1319 nm long-pass) and NdNCs (808 nm excitation, 34 mWcm$^{-2}$, 900 nm long-pass) in the hindlimb of mouse, respectively. The images were taken 1 hour post injection (1 h p. i.) with 10 μl ErNCs and 10 μl NdNCs mixture at the same position. Scale bars: 5 mm. **h**, Comparison on vessel full width at half-maximum width and peak signal to background ratio of the cross-sectional intensity profiles along the dashed line in (**g**).

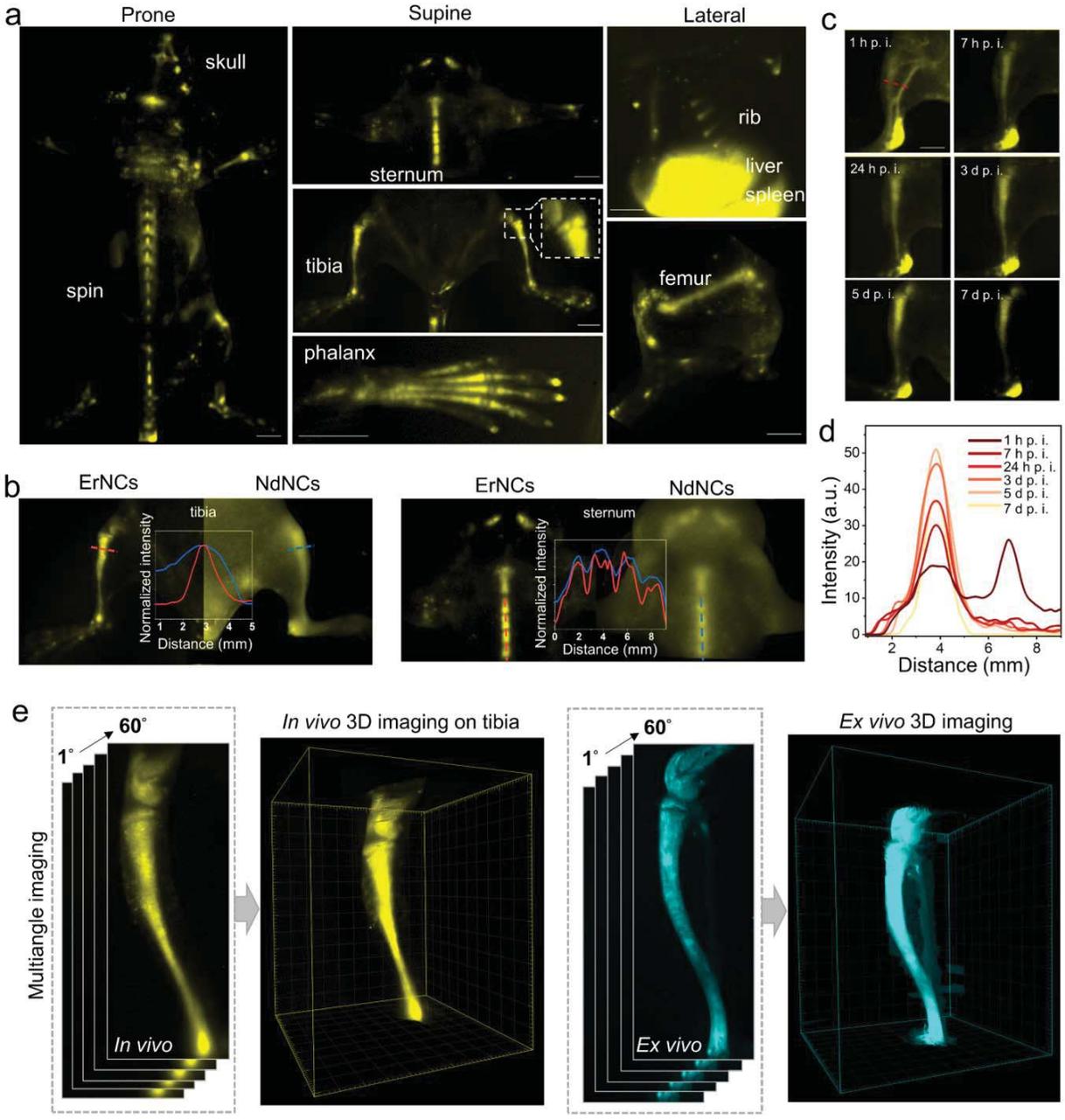

**Figure 2. High-resolution NIR-II imaging of mice bones in both two and three dimensions. a**, NIR-II *in vivo* imaging of murine bones in prone (skull, spin), supine (sternum, tibia, phalanx, subchondral bone in the zoom-in picture), and lateral (rib, femur) postures by the 1550 nm fluorescent ErNCs (980 nm excitation, 38 mWcm⁻², 1319 nm long-pass). Scale bars: 2 mm. **b**, The NIR-II imaging of bone and corresponding cross-sectional intensity profiles by ErNCs (980 nm excitation, 38 mWcm⁻², 1319 nm long-pass) and NdNCs (808 nm excitation, 34 mWcm⁻², 900 nm long-pass), respectively. **c**, *In vivo* NIR-II imaging of mouse tibia by ErNCs (980 nm excitation, 38 mWcm⁻², 1319 nm long-pass, 60 ms) at the different post-injection time points (d: day). Scale bar: 2 mm. **d**, Cross-sectional

intensity profiles along the same position of the time series images represented by the red dash line in (**c**). **e**, The 3D NIR-II imaging on mouse tibia reconstructed by a series of images recorded under different angles (0°-60°) rotated along the central axis of the tibia.

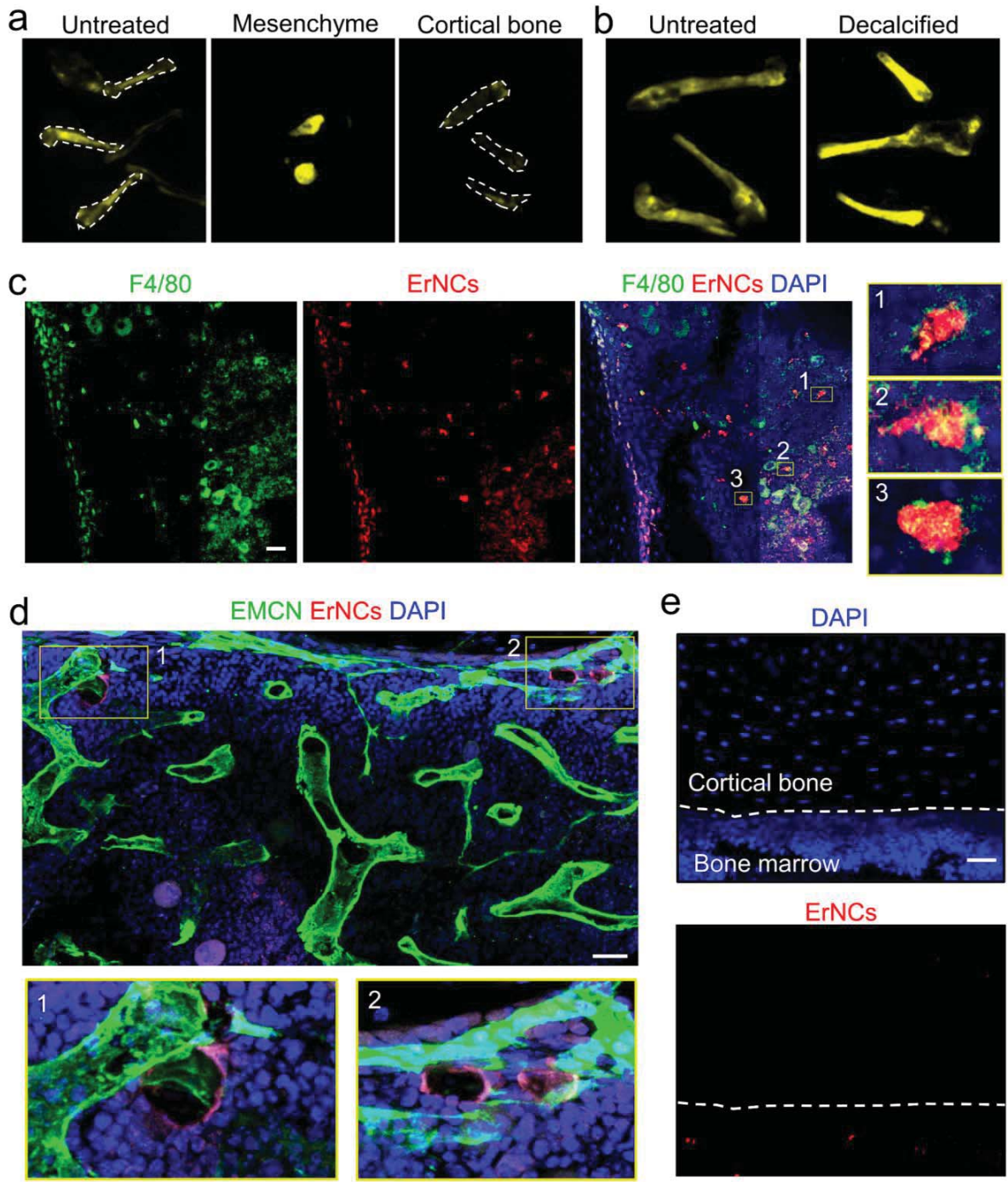

**Figure 3. Cell uptakes of ErNCs in mouse bone marrow. a**, *Ex vivo* NIR-II imaging of the untreated mice femurs, the marrow mesenchyme of the femurs and the femurs without mesenchyme after administration of ErNCs (980 nm excitation, 38 mWcm$^{-2}$, 1319 nm long-pass, 200 ms). **b**, *Ex vivo* NIR-II imaging comparison between the undecalcified bones and the decalcified bones after administration of ErNCs (980 nm excitation, 38 mWcm$^{-2}$, 1319 nm long-pass, 300 ms). **c**, Confocal images of bone marrow in the stained tibia sections collected from a mouse 36 h

after ErNCs@Cy3 injection, including three zoom-in regions of interest. Green channel: F4/80 labeled macrophages. Red channel: ErNCs@Cy3. Blue channel: DAPI labeled cell nucleus. Scale bar: 30 μm. **d**, Confocal images of bone marrow in the stained tibia sections collected from a mouse 1 h post ErNCs@Cy3 injection, including two zoom-in regions of interest. Green channel: Endomucin (EMCN) labeled endothelial cells. Red channel: ErNCs@Cy3. Blue channel: DAPI labeled cell nucleus. Scale bar: 30 μm. **e**, Confocal images of the stained cortical bone area from the same mouse tibia in (**c**). Scale bar: 30 μm.

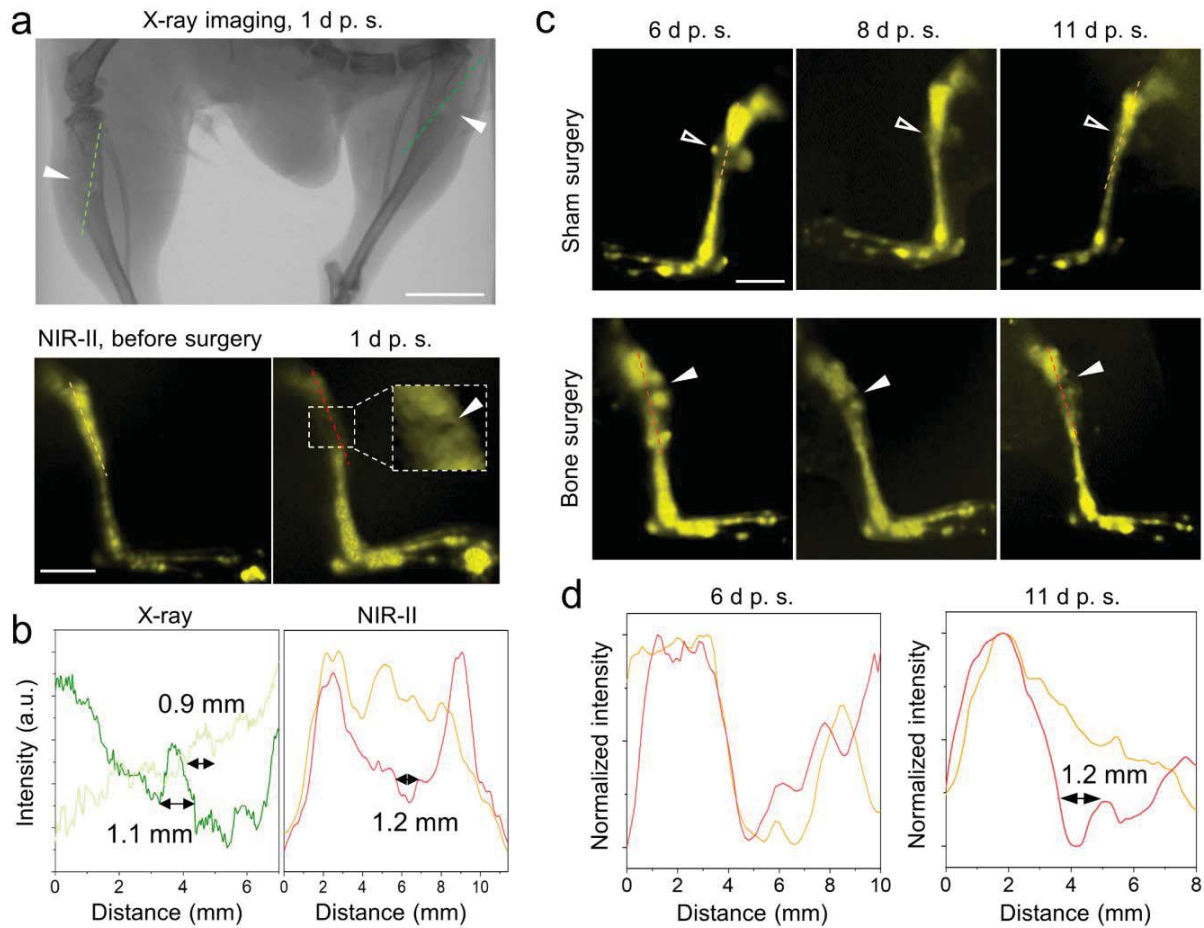

**Figure 4. Long-term NIR-II *in vivo* imaging of bone defects on mice tibias in contrast to X-ray inspection. a**, Comparison between the X-ray and NIR-II *in vivo* imaging on the tibial defects (solid arrowheads). The X-ray imaging was measured 1 d post surgery (p. s.), the NIR-II imaging was taken before and 1 d after surgery, respectively. Scale bars: 5 mm. **b**, The intensity profiles along the cross sections are indicated by the corresponding-colored dash lines in (**a**). The measured sizes of the tibia defects were given. **c**, The NIR-II *in vivo* imaging on the right (sham surgery as control) and left hindlimb (bone surgery) of a mouse on Day 6, 8 and 11 after surgery. The right hindlimb has a soft tissue wound (open arrowheads) from a sham surgery, while the left hindlimb contains both the soft tissue wound and the bone defect (solid arrowheads). Injection of ErNCs was 5 days after surgery via the tail vein. Scale bar: 5 mm. **d**, The intensity profiles along the cross sections are represented by the dash lines in both the sham surgery group (orange) and bone surgery group (red) in (**c**).

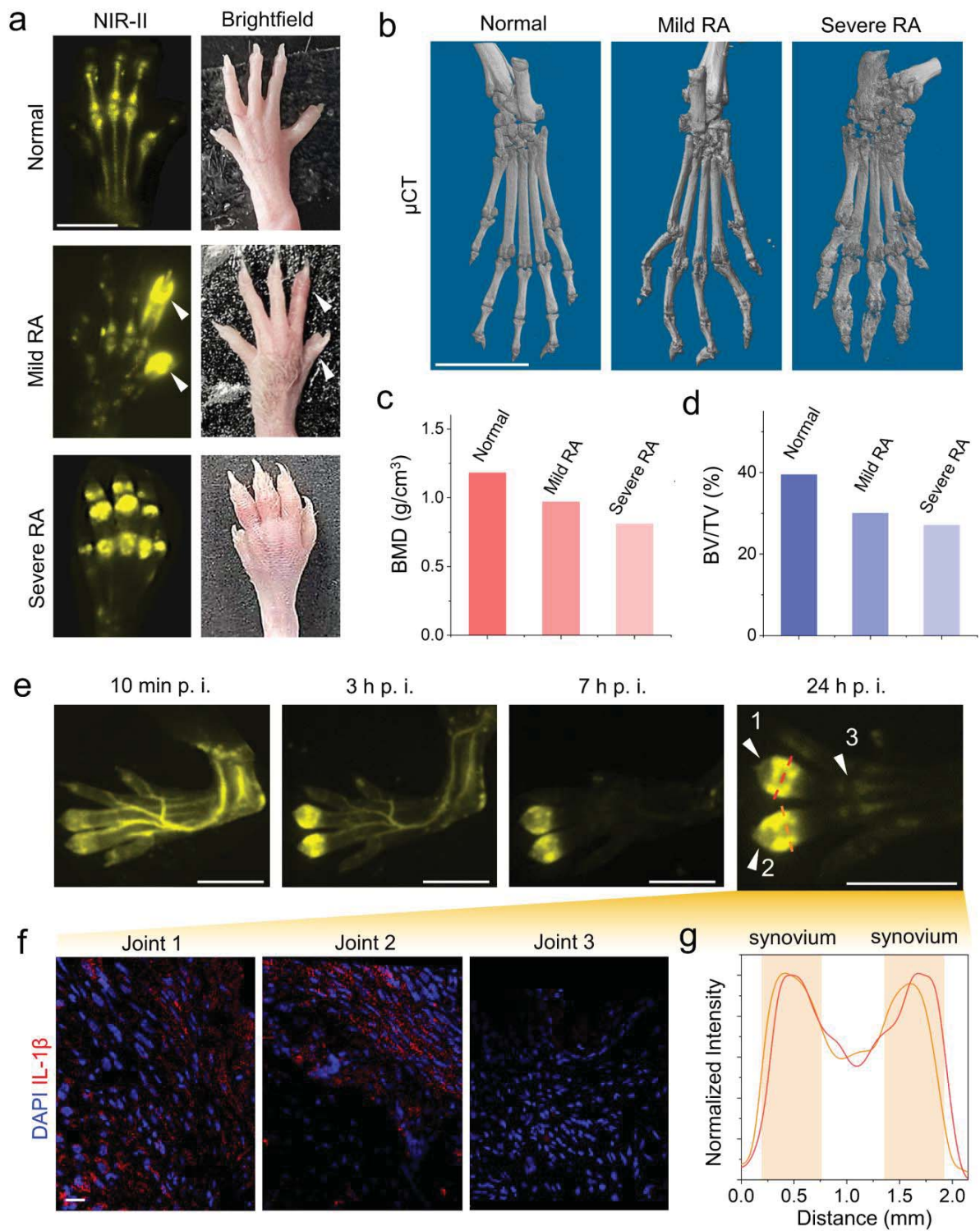

**Figure 5. Early recognition of rheumatoid arthritis (RA) through accurate inflammation NIR-II imaging. a,** The hind paws from a normal mouse, a CIA model mouse with mild RA, and a CIA model mouse with severe RA under NIR-II imaging and bright field, respectively. Scale bars: 5 mm. **b,** Representative μCT images of hind paws

from the normal mouse, CIA mouse with early-stage arthritis, and CIA mouse with late-stage arthritis. Scale bars: 5 mm. **c**, **d**, Quantitative μCT analyses of bone mineral density (BMD) (c) and the bone volume fraction (BV/TV) (d) in toe joints shown in (b), respectively. **e**. Time series NIR-II imaging on the CIA mouse hind paw with early-stage arthritis. Scale bars: 5 mm. **f**, Immunofluorescence analyses of inflammatory factor IL-1β in the synovium tissues from three toe joints (numbered arrowheads) in the hind paw in (e). Scale bar: 20 μm. **g**, The normalized intensity profiles along the cross sections indicated by the dash lines in (e).

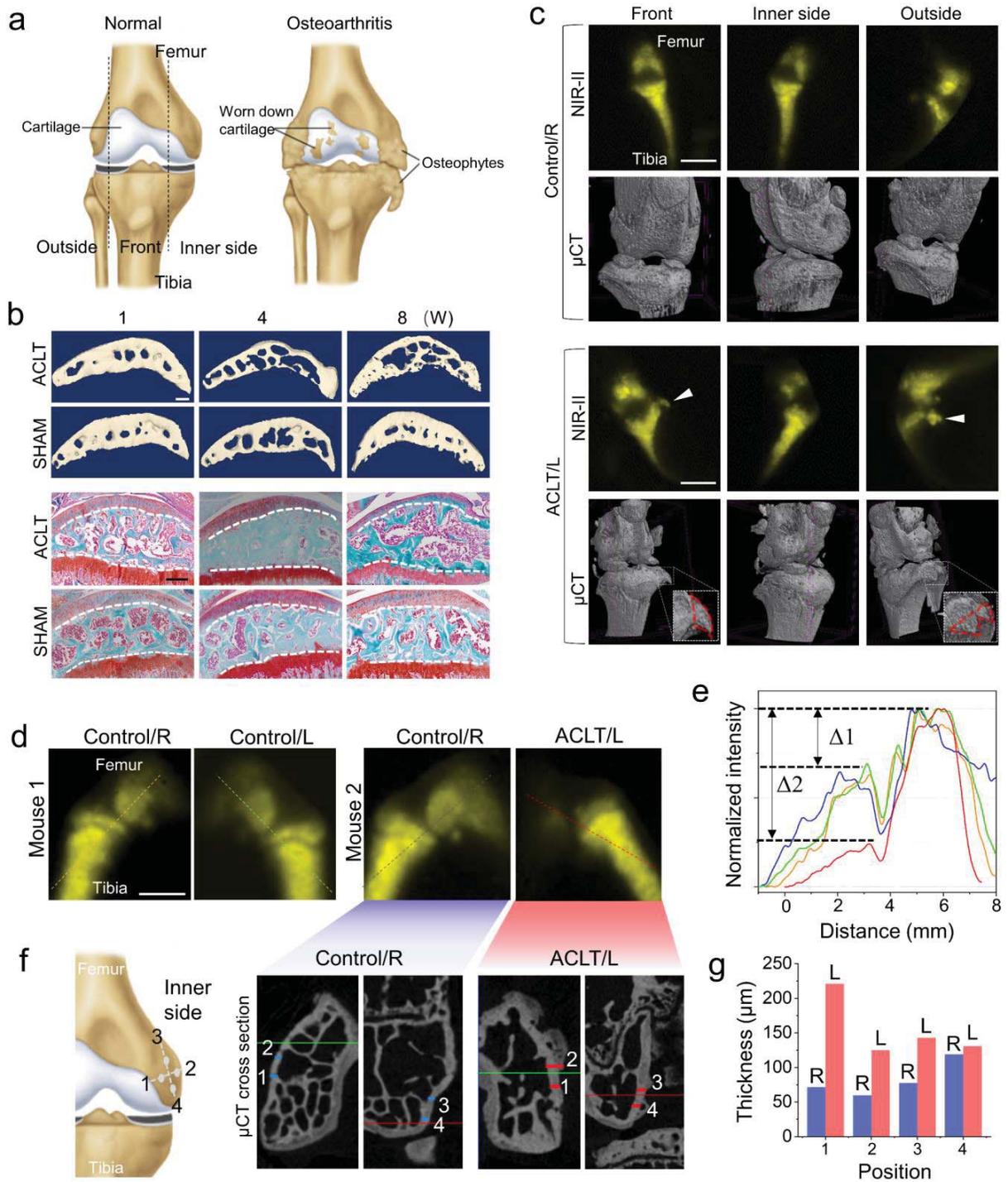

**Figure 6. *In vivo* NIR-II imaging to diagnose osteoarthritis confirmed by μCT. a**, The illustration of osteoarthritis *v.s.* normal arthrosis. **b**, Representative reconstructed μCT images of coronal sections of medial tibia plateaus (first and second rows) and safranin O-fast green staining of tibia plateaus (third and fourth rows), during the time-course of 1, 4, and 8 weeks after ACLT and sham surgery, respectively. Scale bars: 200 μm. **c**, *In vivo* NIR-II imaging and the corresponding reconstructed *ex vivo* µCT images on both hind knee joints of the same mouse, the left knee joint (ACLT/L) had an ACLT surgery 8 weeks ago while the right (Control/R) as the control has no surgery. The solid arrowheads and the red dash lines indicate an osteophyte in the left knee. Scale bars: 5 mm. **d**, *In vivo* NIR-II imaging on the knee joints from two mice, only the left knee of Mouse 2 (ACLT/L) had an ACLT surgery 8 weeks ago, two knees of Mouse 1 and the right knee of Mouse 2 were untreated as control (L: left, R: right). Scale bar: 2 mm. **e**. The normalized intensity profiles along the cross sections are indicated by the dash lines in (**d**). (orange: Right knee of Mouse 1, green: Left knee of Mouse 1, blue: Right knee of Mouse 2, red: Left knee of Mouse 2). **f**, The *ex vivo* µCT cross-section images on the left and right femurs of Mouse 2 in (d). The locations of the cross sections on the femurs are illustrated as well. **g**, Comparison of the cortical bone thicknesses of all positions indicated by the numbered short lines (blue: Right femur; red: Left femurs) in (**f**) from both femurs of Mouse 2 (L: left, R: right).